\documentclass[prd,nofootinbib,floatfix,twocolumn,showpacs]{revtex4}

\usepackage{graphicx,epsfig}
\usepackage{bm}
\usepackage{latexsym,amssymb,amsmath,float}

\def\U{{\cal U}}
\def\V{{\cal V}}

\begin{document}

\title{Voids of dark energy}

\author{Sourish Dutta}
\email{sourish.dutta@case.edu}
\author{Irit Maor}
\email{irit.maor@case.edu}
\affiliation{
CERCA, Department of Physics, Case Western Reserve University,
10900 Euclid Avenue, Cleveland, OH 44106-7079, USA.}

\date{\today}

\begin{abstract}
\noindent
We investigate the clustering properties of a dynamical dark energy component. In a cosmic mix of a pressureless fluid and a light scalar field, we follow the linear evolution of spherical matter perturbations. We find that the scalar field tends to form underdensities in response to the gravitationally collapsing matter. We thoroughly investigate these voids for a variety of initial conditions, explain the physics behind their formation and consider possible observational implications. Detection of dark energy voids will clearly rule out the cosmological constant as the main source of the present acceleration.
\end{abstract}


\maketitle

\section{Introduction}
\label{intro}
Direct \cite{sn} and indirect \cite{nd} evidence of the present
acceleration of the universe is accumulating. Nonetheless, the source of the accelerated expansion is as elusive as ever. In the context of general relativity, one needs to hypothesize a dark energy source with negative pressure to accommodate the acceleration. Alternative ideas include various modifications to gravity. Little is known about the dark energy except that its pressure is negative, and that it accounts for about $70\%$ of the critical density of the universe. Whether the dark energy is the cosmological constant or a dynamically evolving source of energy is a fundamental unanswered question. While the cosmological constant has a fixed ratio of pressure to energy density, $w=p/\rho=-1$, dynamical dark energy (DDE) will in general have a varying equation of state (EOS), $w(z)$. Observing a deviation from $-1$ or a time-evolution in the EOS will be decisive evidence in favor of the existence of DDE. However, there are known degeneracies \cite{pitfalls} which make this task extremely difficult, unless the deviation from a cosmological constant is strong. The current observational limit on the EOS of dark energy is roughly $w\approx-1\pm 0.1$ at the $1\sigma$ level \cite{sn}, which is consistent with a cosmological constant.
Future experiments hold out the possibility of pinning down this limit by maybe a factor of $10$. For a recent review, see \cite{review}.

The effect of the cosmological constant and DDE on the expansion rate
can be identical, and there is a need for probes that go beyond the background to distinguish between the two. An example of such a probe is structure formation. There are numerous works exploring the formation of structure in the presence of homogeneous dark energy \cite{hdde}. An exciting and somewhat controversial possible difference between the cosmological constant and DDE is their clustering behavior. While the cosmological constant is exactly homogeneous on all scales, DDE is expected to be not perfectly homogeneous \cite{nh}, and the implications of this on the CMB are well known \cite{cmbnh}. However, it is usually assumed that the clustering of DDE is negligible on scales less than $100~{\rm{Mpc}}$. Whether small perturbations in DDE can be neglected is debatable, and a deeper understanding of the DDE inhomogeneous dynamics is clearly needed.

Several recent works have explored the consequences of DDE clustering on scales shorter than $100~{\rm{Mpc}}$. Some have adopted a phenomenological approach, parameterizing the clustering degree of DDE \cite{parametric}. These works point out potential observables of DDE clustering, justifying further investigation. Works which attempt a more fundamental treatment are mostly in the context of coupled dark energy \cite{cde}, or other non-trivial models of DDE \cite{other}, as clustering is most probable in such theories. However, less attention has been given to the clustering in simpler models of DDE.

In the present work, we aim to further explore the inhomogeneous behavior of DDE. Our approach is straightforward: starting with a gravitational action which includes matter and DDE, we numerically follow the linear evolution of spherical perturbations of matter and the DDE response to these perturbations. For the sake of simplicity, our model for the DDE is a light scalar field, which is not explicitly coupled to the matter density. As the only coupling between the DDE and the matter is gravitational, our results are conservative in the sense that any model more complicated will exhibit stronger DDE perturbations than shown here. The striking feature that emerges from our calculation is that in the vicinity of collapsing matter, the DDE develops a spatial profile and tends to form voids. The mechanism that allows the void to form is that although initially the field's evolution is friction dominated due to the cosmic expansion, the collapse of matter slows down the local expansion. This allows the field to locally roll down and lose energy, creating the void. The presence of the matter perturbation is necessary to trigger this mechanism.

The plan of the paper is as follows: in section \ref{The Model} we describe our model in detail. In section \ref{results} we present our results. We discuss our results and state our conclusions in sections \ref{Discussion} and \ref{Conclusions} respectively.


\section{The model}
\label{The Model}

We are interested in spherical perturbations around a flat FRW universe. The most general line element in comoving coordinates is then
\begin{equation}
\label{metric}
	ds^{2}=dt^2-\U(t,r)dr^2-\V(t,r)\left(
	d\theta^2+\sin^{2}\theta d\varphi^2 \right) ~,
\end{equation}
where $\U(t,r)$ and $\V(t,r)$ are general functions \cite{weinberg}.

We take a cosmic mix of non-relativistic matter and a DDE component as the energy source. The matter component is described by a perfect and pressureless fluid, with an energy-momentum tensor given by
\begin{equation}
	\label{T_munu matter}
	T_{\mu\nu}(m)=\text{diag}\left(\rho,0,0,0\right)~,
\end{equation}
where $\rho$ is the energy density of matter.

We model the DDE with a classical scalar field $\phi$ with a Lagrangian $\cal L$  given by
\begin{eqnarray}
	\label{phi lagrangian}
	{\cal L} &=& \frac{1}{2}\left(\partial_\mu\phi\right)^2-V(\phi) ~,
\end{eqnarray}
and an energy-momentum tensor given by
\begin{equation}
\label{T_munu phi}
	T_{\mu\nu}(\phi)=\partial_{\mu}\phi\partial_{\nu}\phi-
	g_{\mu\nu}{\cal L} ~.
\end{equation}

The EOS of the DDE $w$ is defined as
\begin{equation}
	\label{w_exact}
	w=\frac{p_{\phi}}{\rho_{\phi}},
\end{equation}
with the energy density $\rho_{\phi}$ and the pressure $p_{\phi}$ are read off the energy momentum tensor, $T_{00}(\phi)$ and $-g^{ij} T_{ij}(\phi)/3$ respectively. \\

It is convenient to rewrite Einstein's equations in the following way,
\begin{equation}
	\label{gurudev}	
R_{\mu\nu}=K\left(T_{\mu\nu}-\frac{1}{2}g_{\mu\nu}T^{\alpha}~_{\alpha}\right)
~.
\end{equation}
where $R_{\mu\nu}$ is the Reimann tensor, and $K=8\pi G $. As there is no explicit interaction between the matter and the DDE, energy conservation applies to each separately,
\begin{equation}
	\label{Energy Conservation}
	\nabla_{\mu}T^{\mu\nu}(m)=0 ~,~~
	\nabla_{\mu}T^{\mu\nu}(\phi)=0 ~.
\end{equation}

The time-evolution of the system is given by the following equations (where dots denote time-derivatives and primes denote derivatives with respect to the radial coordinate, except for $V'(\phi) \equiv \frac{d V}{d \phi}$):
\begin{eqnarray}
	\frac{1}{2}\frac{\ddot{\U}}{\U}+
	\frac{1}{2}\frac{\dot{\U}}{\U}\frac{\dot{\V}}{\V}-
	\frac{1}{4}\frac{\dot{\U}^2}{\U^2}+
	\frac{1}{\U}\left(\frac{1}{2}\frac{\U'}{\U}\frac{\V'}{\V}+
	\frac{1}{2}\frac{\V'^2}{\V^2}-
	\frac{\V''}{\V}\right) &&~~ \nonumber \\
	\label{U} \qquad
	-K\left(\frac{1}{2}\rho+
	V(\phi)+
	\frac{1}{\U}\phi'^2\right)=0 ~~&&~~~ \\
	\frac{1}{2}\frac{\ddot{\V}}{\V}+
	\frac{1}{4}\frac{\dot{\U}}{\U}\frac{\dot{\V}}{\V}+
	\frac{1}{\V}+
	\frac{1}{2\U}\left(\frac{1}{2}\frac{\U'}{\U}\frac{\V'}{\V}-
	\frac{\V''}{\V}\right) ~~&&~~ \nonumber \\
	\label{V}\qquad
	-K\left(\frac{1}{2}\rho+
	V(\phi)\right)=0 ~~&&~~~ \\
	\label{rho}\dot{\rho}+
	\left(\frac{\dot{\V}}{\V}+
	\frac{1}{2}\frac{\dot{\U}}{\U}\right)\rho=0 ~~&&~~~ \\
	\ddot{\phi}+\left(\frac{\dot{\V}}{\V}+
	\frac{1}{2}\frac{\dot{\U}}{\U}\right)\dot{\phi}+
	V'(\phi)
	\qquad\qquad\qquad ~~&&~~~ \nonumber \\
	\label{phi}
	-\frac{1}{\U}\left( \phi''+
    \Big(\frac{\V'}{\V}-\frac{1}{2}\frac{\U'}{\U}\Big)
    \phi' \right)=0~. &&~~~~~~
\end{eqnarray}
These are subject to the following constraint equations:
\begin{eqnarray}
	\frac{1}{\V}+
	\frac{1}{2}\frac{\dot{\U}}{\U}\frac{\dot{\V}}{\V}+
	\frac{1}{4}\frac{\dot{\V}^2}{\V^2}+
	\frac{1}{\U}\left( \frac{1}{2}\frac{\U'}{\U}\frac{\V'}{\V}+
	\frac{1}{4}\frac{\V'^2}{\V^2}-
	\frac{\V''}{\V}\right) && \nonumber \\
	\label{friedmann}\qquad
	-K\left(\rho+V(\phi)+
	\frac{1}{2}\dot{\phi}^2+
	\frac{1}{2\U}\phi'^2 \right)=0 ~~&&~~~ \\
	\label{diag}\frac{1}{2}\frac{\dot{\U}}{\U}\frac{\V'}{\V}+
	\frac{1}{2}\frac{\dot{\V}}{\V}\frac{\V'}{\V}-
	\frac{\dot{\V}'}{\V}-
	K \dot{\phi}\phi'= 0~. &&~~~~~~
\end{eqnarray}


\subsection{Linearization}
\label{Linearization}

We now proceed to separate our variables to a homogeneous background and a time and space-dependent perturbation, which we will then linearize.
Working in the synchronous gauge \cite{gauge}, we redefine the metric functions $U$ and $V$ as follows:
\begin{eqnarray}
	\U(t,r) &=& a(t)^2 e^{2\zeta(t,r)} \nonumber \\
	\V(t,r) &=& r^2 a(t)^2 e^{2\psi(t,r)} ~. \nonumber \\
\end{eqnarray}
Here $a(t)$ is the scale factor of the spatially homogenous and flat background, and $\zeta(t,r)$ and $\psi(t,r)$ are the deviations. We introduce a perturbation around a homogeneous background also in the matter and the DDE,
\begin{eqnarray}
    \rho(t,r)&=&\rho(t)+\delta\rho(t,r) \nonumber \\
    \phi(t,r)&=&\phi(t)+\delta\phi(t,r) \nonumber \\
    V(\phi+\delta\phi) &=& V(\phi)+\delta V(\phi,\delta\phi) \nonumber ~.
\end{eqnarray}

The zeroth order of equations \eqref{U} - \eqref{diag} gives
\begin{eqnarray}
    \label{fried0}3H^{2}
	-K\left[\rho+V+
	\frac{1}{2}\dot{\phi}^2\right]&=&0   \\
	\label{H0} \dot{H}+3H^2-K\left(\frac{1}{2}\rho+
	V\right)&=&0  \\
    \label{rho0} \dot{\rho}+3H\rho&=&0  \\
    \label{phi0} \ddot{\phi}+3H\dot{\phi}+V'&=&0 ~,
\end{eqnarray}
where $H=\dot a/a$ is the Hubble function. \\
	
To linear order, the evolution equations \eqref{U}-\eqref{phi} give
\begin{eqnarray}
	\ddot{\zeta}+4H\dot{\zeta}+2H\dot{\psi}+
	\frac{2}{a^2}\left( \frac{\zeta'}{r}-
	\frac{2\psi'}{r}-\psi''\right)
	\qquad\qquad ~~ && \nonumber \\
	\label{zeta}-K\left(\frac{1}{2}\delta\rho+
	\delta V\right)=0 \qquad ~~ && \\
	\ddot{\psi}+5H\dot{\psi}+H\dot{\zeta}+
	\frac{1}{a^2}\left(\frac{2\zeta}{r^2}-
	\frac{2\psi}{r^2}+\frac{\zeta'}{r}-
	\frac{4\psi'}{r}-\psi''\right) \nonumber \\
	\label{psi}-K\left(\frac{1}{2}\delta\rho+
	\delta V\right)=0 \qquad ~~ && \\
	\label{drho}\delta\dot{\rho}+3H\delta{\rho}+
	\rho\left(\dot{\zeta}+2\dot{\psi}\right)=0 \qquad ~~ && \\
	\delta\ddot{\phi}+3H\delta\dot{\phi}+
	\delta V'
	\qquad\qquad\qquad\qquad\qquad\qquad ~~ && \nonumber \\
	\label{dphi} +\left(\dot{\zeta}+
	2\dot{\psi}\right)\dot{\phi}-
	\frac{1}{a^2}\left(\delta\phi''+
	\frac{2}{r}\delta\phi'\right)=0  ~,\qquad &&
\end{eqnarray}
and the constraint equations \eqref{friedmann}-\eqref{diag} reduce to
\begin{eqnarray}
	2H\dot{\zeta}+4H\dot{\psi}+
	\frac{2}{a^2}\left(\frac{\zeta}{r^2}-
	\frac{\psi}{r^2}+\frac{2\zeta'}{r}-
	\frac{6\psi'}{r}-\psi''\right)\qquad ~~ && \nonumber \\
	\label{fried}-K\left(\delta\rho+\delta V+
	\dot{\phi}\delta\dot{\phi}\right)=0\qquad ~~ && \\
	\label{cons} \frac{2}{r}\left(\dot{\psi}+
	\dot{\zeta}\right)-K\dot{\phi}\delta\phi'+
	2\dot{\psi}'=0 ~. \qquad &&
\end{eqnarray}

Combining equations \eqref{zeta}, \eqref{psi} and \eqref{fried} gives
\begin{eqnarray}
    \left(
	\ddot{\zeta}+2\ddot{\psi} \right)
    +2H\left(\dot{\zeta}+
	2\dot{\psi}\right)
	\qquad\qquad\qquad\qquad\qquad ~~ && \nonumber \\
	\label{manip} +K\left(\delta\rho-\delta V+
	2\dot{\phi}\delta\dot{\phi}\right)=0 ~. \quad
\end{eqnarray}

The only combination which is relevant to the equations of motion \eqref{drho} and \eqref{dphi} is $\chi\equiv\dot{\zeta}+2\dot{\psi}$. Comparing \eqref{rho0} and \eqref{drho} it is clear that $\chi$ can be thought of as $3 \delta H$, and therefore characterizes the spatial profile of the Hubble function.
At the cost of losing some information about the metric, we can reduce the number of our variables and equations from $4$ to $3$ by solving  for $\chi$ instead of  for $\zeta$ and $\psi$. Equations ~\eqref{drho}, \eqref{dphi} and \eqref{manip} yield
\begin{eqnarray}
    \label{drho1}\delta\dot{\rho}+3H\delta{\rho}+\rho\chi=0 ~~ && \\
    \label{dphi1}\delta\ddot{\phi}+3H\delta\dot{\phi}+\delta V'+
    \chi\dot{\phi}-\frac{1}{a^2}\nabla^2\left(\delta\phi\right)=0 ~~ && \\
    \label{manip1}\dot{\chi}+2H\chi+K\left(\frac{1}{2}\delta\rho-
    \delta V+2\dot{\phi}\delta\dot{\phi}\right)=0 ~. &&
\end{eqnarray}
Finally, by Fourier-transforming $\delta\rho(t,r)$, $\delta\phi(t,r)$ and $\chi(t,r)$ into $\delta\rho_k(t,k)$, $\delta\phi_k(t,k)$ and $\chi_k(t,k)$ respectively, equations \eqref{drho1}-\eqref{manip1} can be written as a set of ordinary differential equations:
\begin{eqnarray}
    \label{drho1k}\delta\dot{\rho_k}+3H\delta{\rho_k}+\rho\chi_k=0 ~~&& \\
    \label{dphi1k}\delta\ddot{\phi_k}+3H\delta\dot{\phi_k}+
    \left(V''+\frac{k^2}{a^2}\right)\delta\phi_k+\dot{\phi}\chi_k=0 ~~&& \\
    \label{manip1k}\dot{\chi_k}+2H\chi_k+K\left(\frac{1}{2}\delta\rho_k
    -V'\delta\phi_k+2\dot{\phi_k}\delta\dot{\phi_k}\right)=0 ~, &&
\end{eqnarray}
where we have used the fact that to linear order, $\delta V=V'\delta\phi$ and $\delta V'=V''\delta\phi$.

\subsection{Potential}

Observationally distinguishing between various potentials of DDE is a formidable task \cite{potential}, and a careful analysis of the growth of structure in various potentials might prove a useful tool. Our present goal though is to trace generic properties of DDE. Accordingly, we choose to work with a simple mass potential,
\begin{eqnarray}
 V(\phi) &=&\frac{1}{2}m^2\phi^2 ~.
 \label{vmass}
\end{eqnarray}
We take the mass scale comparable to the present Hubble scale, $m_{\phi}/H_0\sim 1$. The light mass assures a slow roll behavior, which will provide accelerated cosmic expansion. Unless noted otherwise, the figures presented here refer to this mass potential.

In order to verify the generality of our results, we repeated the analysis for a more realistic potential - the double exponential \cite{barreiro},
\begin{equation}
 V(\phi)=V_{0}\left(e^{\sqrt{K}\alpha\phi}+e^{\sqrt{K}\beta\phi}\right) ~,
 \label{v2exp}
\end{equation}
with $\alpha=20.1$ and $\beta=0.5$. As we later show, the resulting behavior for the two potentials is qualitatively the same.

\subsection{Initial conditions}

We want to study how the DDE reacts to the clustering of matter. Thus our initial conditions are of perturbed matter and homogeneous DDE. The matter perturbation is taken as a spherical gaussian,
\begin{equation}
	\delta_m(t_i,r)\equiv\delta\rho_m(t_i,r)/\rho_m(t_i)=
	A\exp\left(-r^2/\sigma^2\right) ~,
\end{equation}
and the DDE is taken to be initially homogeneous, $\delta\phi=0$. A non-homogeneous evolution for the dark energy is nonetheless allowed. To ensure that the matter perturbation has no peculiar velocity, we choose the initial condition for the metric variable as $\chi=0$. From equation \eqref{drho1}, we find that this amounts to the statement that initially, $\delta\rho\propto a^{-3}$, {\it{i.e.}} the matter particles making up the perturbation are being simply carried along with the Hubble expansion (or in other words, they are at rest in a comoving frame). However, we have verified that choosing different initial values of $\chi$ does not qualitatively affect our results.

The numerical calculation begins at a redshift of $z=35$ with a perturbation amplitude of $A=0.1$, and is run to the present, $z=0$. We focus on relatively short-scale perturbations, $\sigma H_i \sim 0.01$. With this choice of redshift span, we aim to explore the full linear range of the matter perturbation, while also allowing for a period of DDE dominance.

The initial values of the background variables $\rho_i$, $\phi_i$ and $H_i$ are chosen such that their present values (denoted with a $0$ subscript) converge to $\Omega_{m,0}=0.3$, $\Omega_{\phi,0}=0.7$, and a normalized Hubble value, $H_0=1$. This is done with the help of a root-finding algorithm. We take $\dot\phi_i=0$, which means that the initial state of the scalar field is homogeneous and with equation of state $w=-1$, similar to the cosmological constant.

We now have the layout to numerically evolve equations \eqref{drho1k}-\eqref{manip1k}. The solutions for $\delta\rho_k$, $\delta\phi_k$ and $\chi_k$ are then Fourier transformed back to real space.

\section{Results}
\label{results}

\subsection{Density contrasts}

\begin{figure}
	\epsfig{file=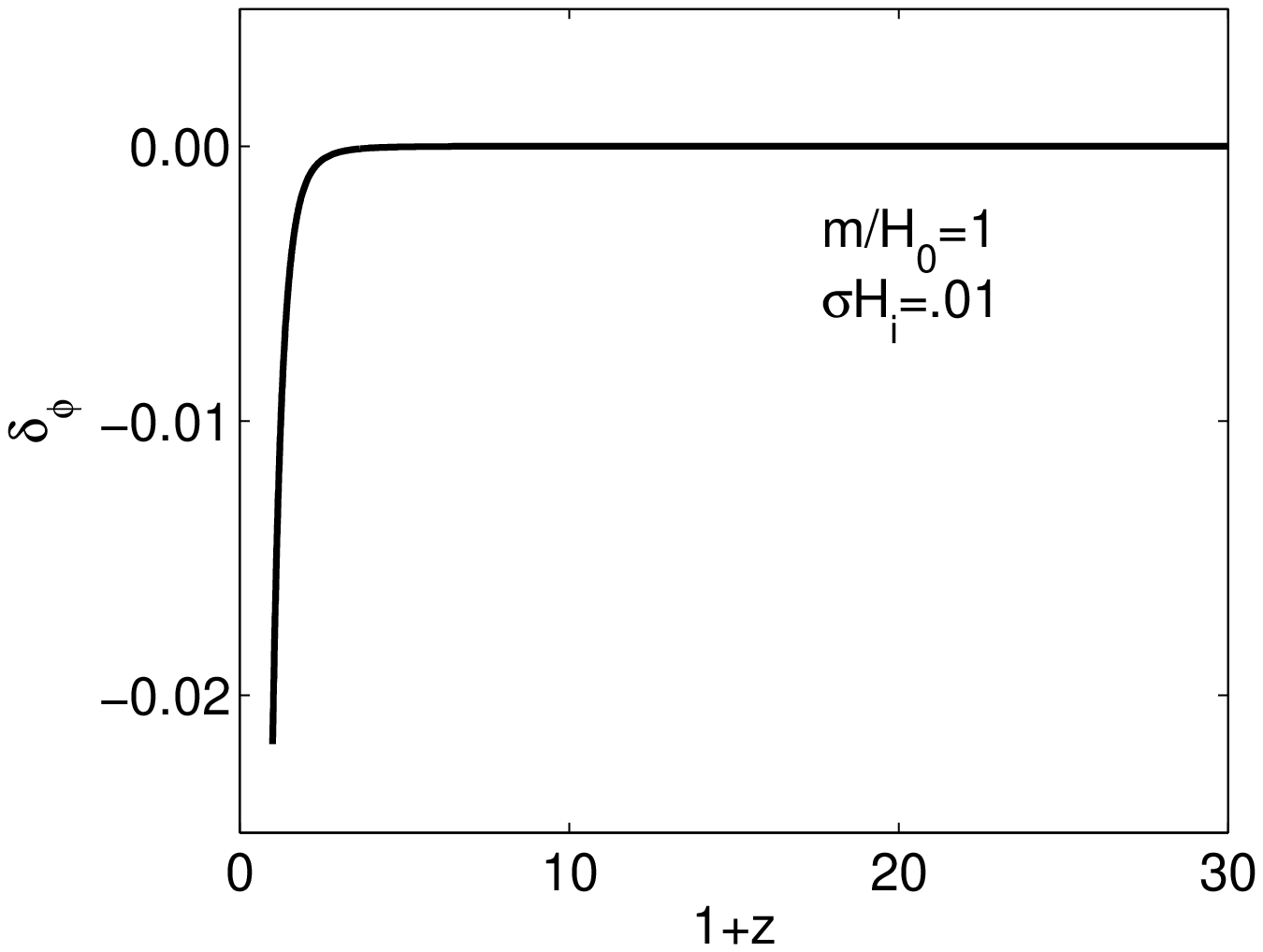,height=55mm}
	\caption{\label{void_demo1} The evolution of the DDE overdensity
	$\delta_{\phi}$ at the center of the matter perturbation,
	$r=0$, with redshift $(1+z)$.
    The scale of the perturbation is $\sigma H_{i}=0.01$, and
    the mass is $m/H_0=1$. Initially homogenous, the DDE develops an
    underdensity at late times in response to the matter perturbation.
}
%
	\epsfig{file=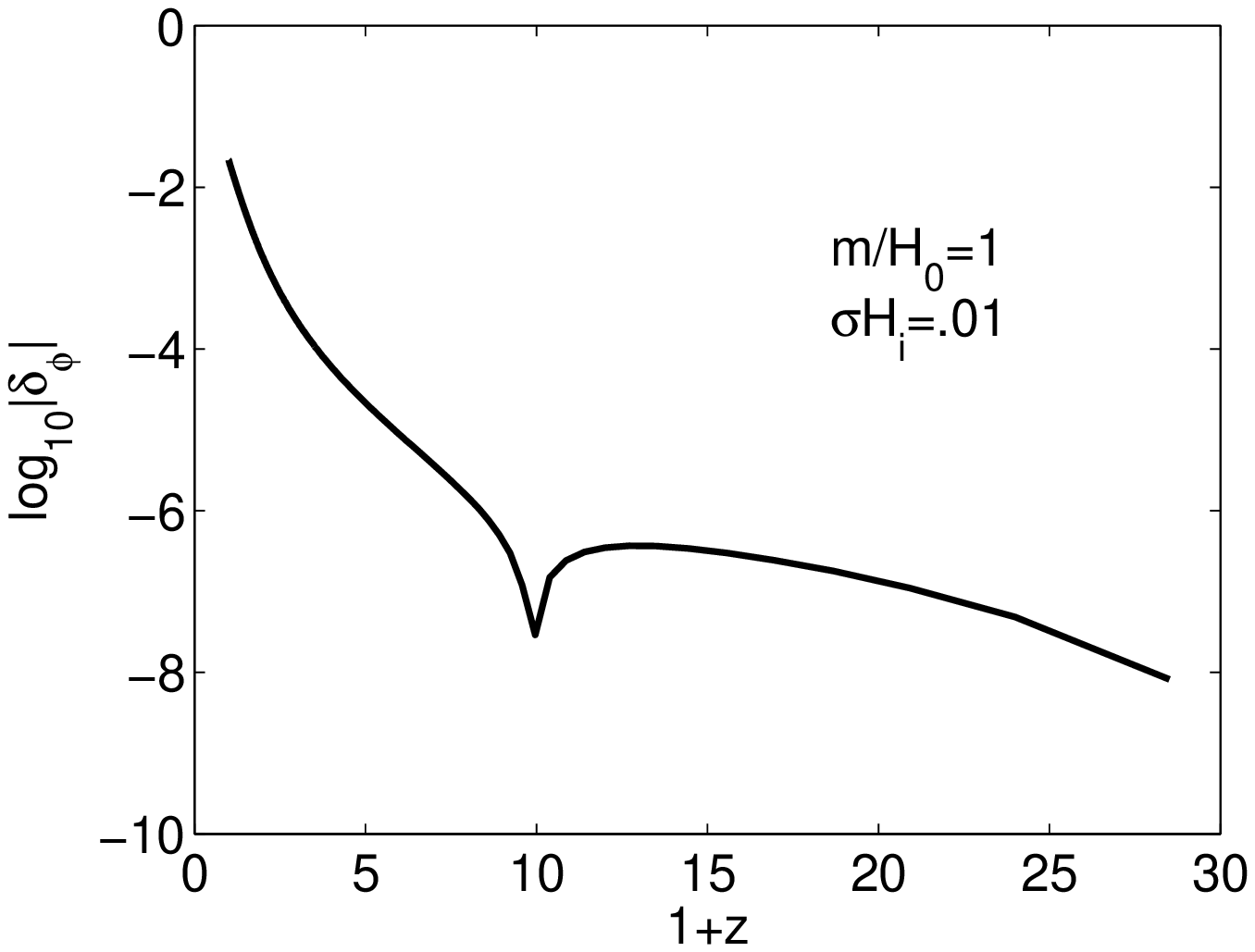,height=55mm}
	\caption{\label{void_demo2} Same as Figure \ref{void_demo1},
    with the y-axis on a logarithmic scale. The DDE
    tends to cluster initially, but eventually forms a void.
}
\end{figure}

The results shown in this section are for the mass potential, equation \eqref{vmass}, unless specifically noted otherwise.

The numerical run begins at a redshift of $z=35$, when the DDE is subdominant. The DDE remains subdominant through most of the growth time of the perturbation as well, and accordingly, we expect the matter density contrast $\delta_{m}\equiv\delta\rho_{m}/\rho_{m}$ to grow as the scale factor $a$. This is indeed confirmed by our results.

The most striking result that emerges from our calculation is that in the vicinity of collapsing matter, the DDE tends to form voids, or growing regions of under-density. The anti-correlation between the perturbations of the matter and the DDE was noted in \cite{cmbnh} and in \cite{nunes} for particular cases. Figure \ref{void_demo1} plots the DDE density contrast $\delta_{\phi}\equiv\delta\rho_{\phi}/\rho_{\phi}$ at the center of the matter perturbation, $r=0$, against the redshift $(1+z)$. We find that the amplitude of the perturbation grows sharply at late times.

Figure \ref{void_demo2} shows the growth of the absolute value of the DDE perturbation with redshift on a logarithmic scale. This plot reveals another interesting effect: the initial response of the DDE to the gravitationally collapsing matter is a very weak tendency to collapse. The collapsing phase, however, is extremely short-lived (an ${\cal{O}}[10^{-1}]$ fraction of the total time of the run). The kink in the figure is the cross-over from a positive to negative perturbation.

\begin{figure}
	\epsfig{file=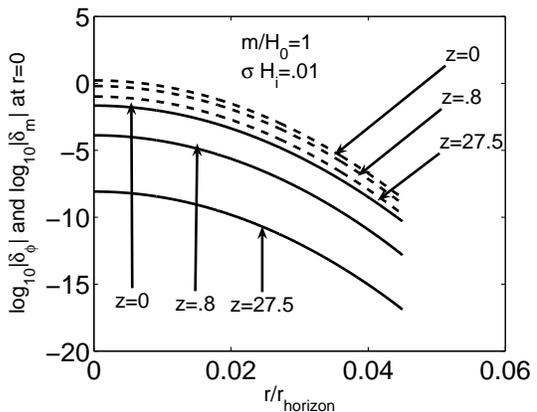,height=55mm}
	\caption{\label{profiles} Logarithmic profiles of the matter
    density contrast $\log_{10}\vert\delta_m\vert$ and the dark
    energy density contrast $\log_{10}\vert\delta_\phi\vert$, at
    three different redshifts. Solid lines denote the DDE
    profiles, and dotted lines denote the matter profiles.}
\end{figure}

We next look at the spatial profiles of the perturbations. Figure \ref{profiles} shows the profiles of both the $\delta_m$ and $\delta_{\phi}$ at an early stage of the run ($z=27.5$), at an intermediate redshift ($z=0.8$) and at the final stage  ($z=0$). The x-axis shows the physical scale as a fraction of the horizon size. The amplitude of the perturbations is shown on a logarithmic scale, and it is worth noting the change in gap between the two scales. The growth rate of $\delta_{\phi}$ is significantly faster than that of $\delta_m$, so that the two amplitudes are almost comparable at late times. This suggests that a calculation of the non-linear regime might reveal interesting behavior of the DDE.

\begin{figure}
    \epsfig{file=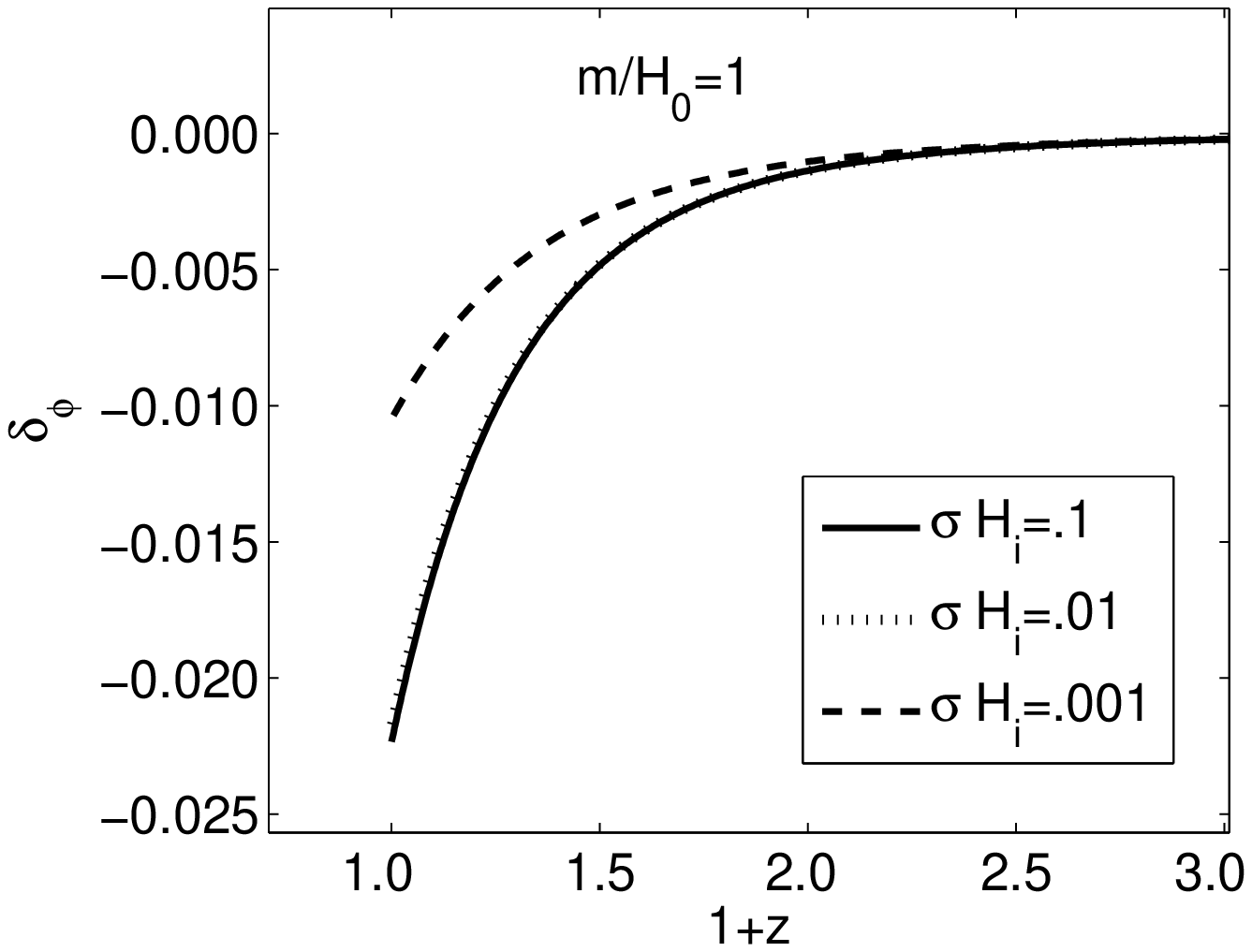,height=55mm}
	\caption{\label{sigma_effect1} DDE density contrast
    $\delta_{\phi}$ at the center of the matter perturbation
    $r=0$ as a function of the redshift $(1+z)$ for fixed mass
    $m/H_0=1$ and different initial matter perturbations' widths.
    The larger the initial matter perturbation, the
    stronger is the void.
    The curves of $\sigma H_i=0.01$ (dashed) and $0.1$ (solid) almost overlap.
    The figure zooms on late times, $z<3$.
}
%
	\epsfig{file=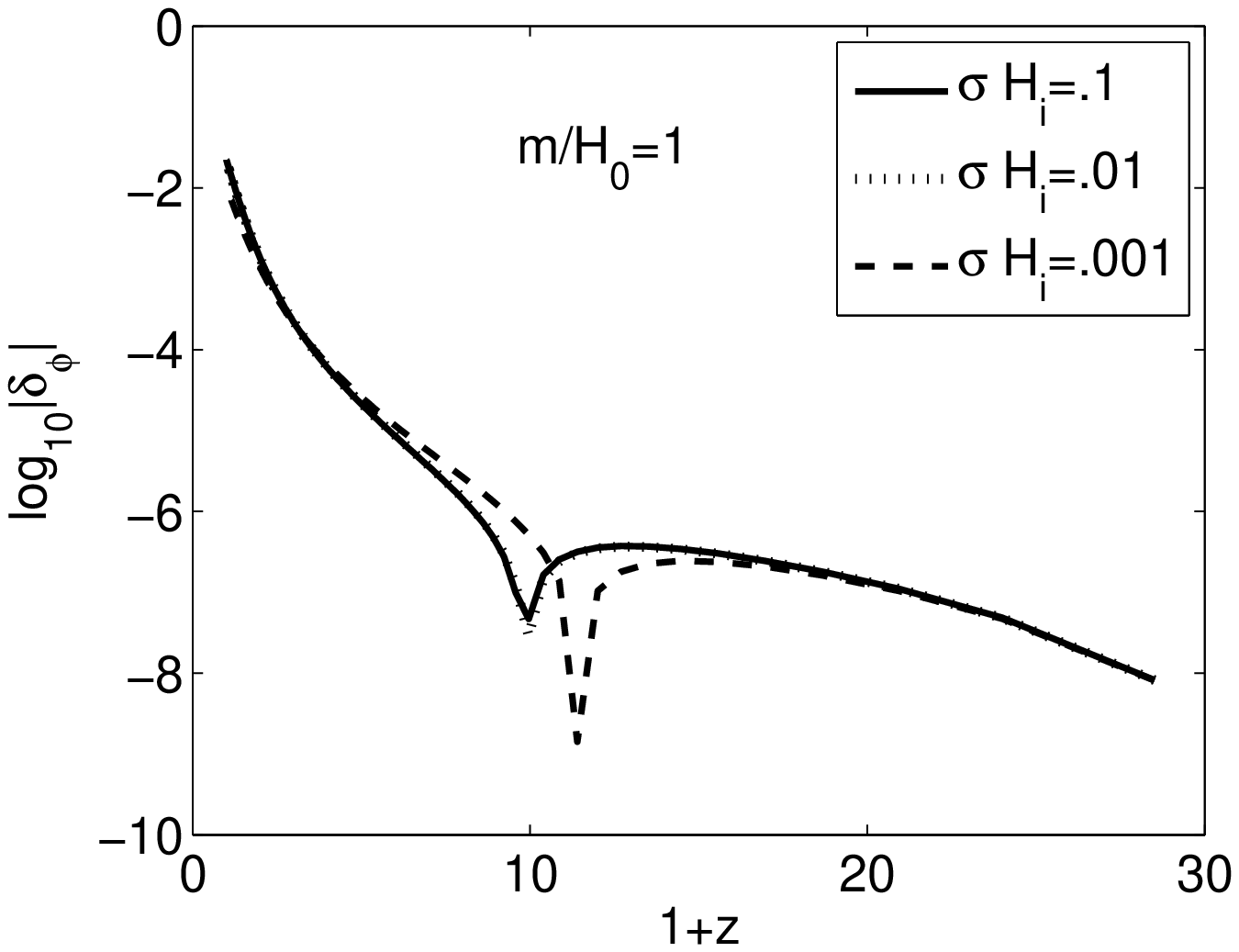,height=55mm}
	\caption{\label{sigma_effect2}
    Same as Figure \ref{sigma_effect1} with the y-axis on a
    logarithmic scale. The shorter scales start evolving at
    later times than the longer scales, but their evolution
    is faster.
    The curves of $\sigma H_i=0.01$ (dashed) and $0.1$ (solid) almost overlap.
}
\end{figure}

Figures \ref{sigma_effect1} and \ref{sigma_effect2} show how the growth of the DDE perturbation is affected by the initial width of the matter perturbation. The figure confirms the sensitivity of DDE perturbations to the scale, and as expected, shorter scales exhibit a suppressed behavior. Nonetheless, a possibly significant amplitude can be found on relevant scales. For instance, our local supercluster has diameter of order $0.01H_{0}^{-1}$ and an overdensity of about $\delta_m\approx 1.2$. Our runs of $\sigma H_i=0.01$ end with an overdensity of $\delta_m\approx 1.6$, but roughly they can be taken as a measure of what we should expect on these scales.
Also, one must remember that this analysis only captures the physics of the linear regime. Given the sharp increase in the void amplitude at later times, it is not unreasonable to expect interesting effects in the strongly non-linear regime.

\begin{figure}
    \epsfig{file=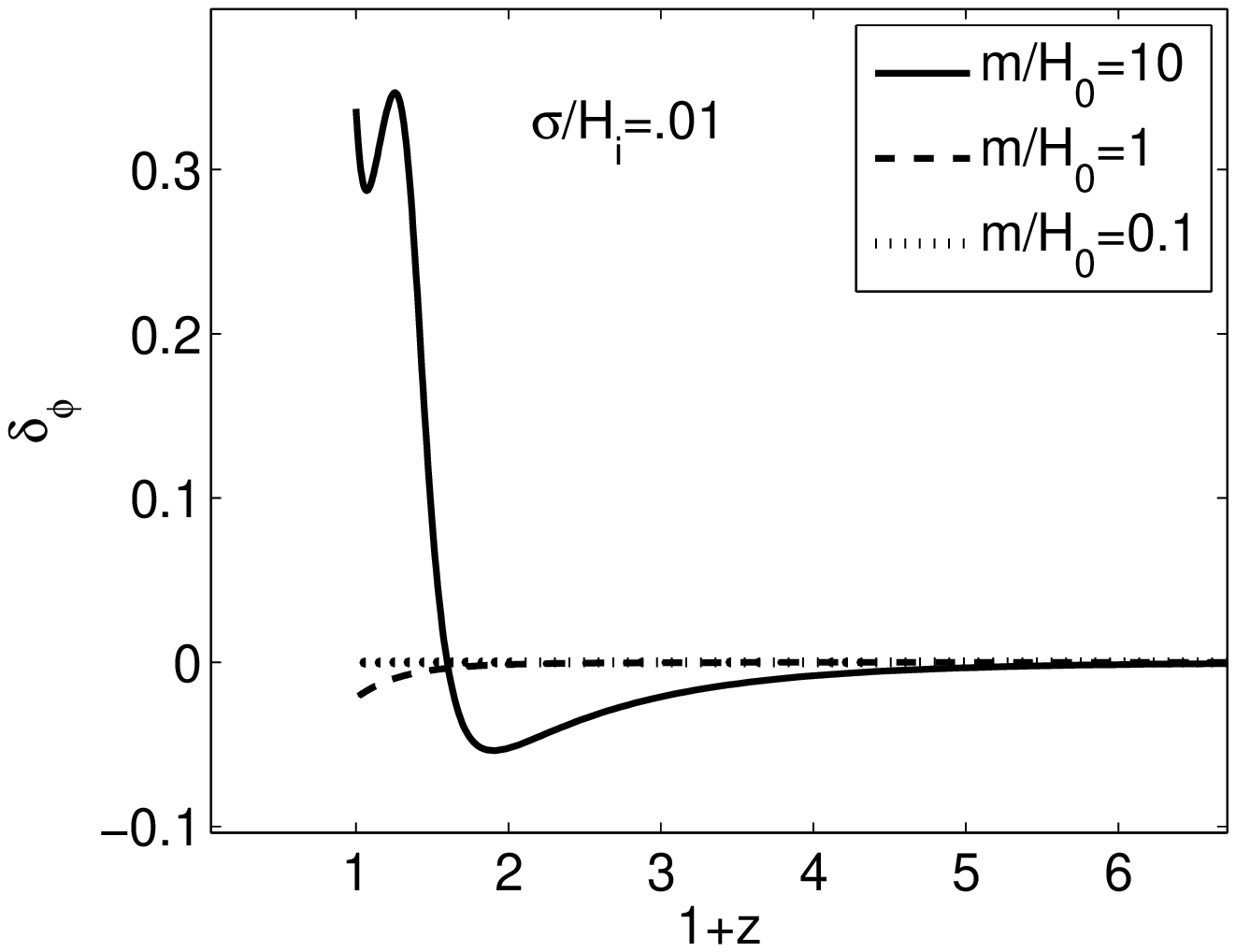,height=55mm}
	\caption{\label{mass_effect1} The DDE density contrast $\delta_{\phi}$
    at the center of the matter perturbation,
    $r=0$, against redshift $(1+z)$ for
    $\sigma H_i=.01$ and three different masses.
    The figure zooms on late times, $z<7$.
}
%
	\epsfig{file=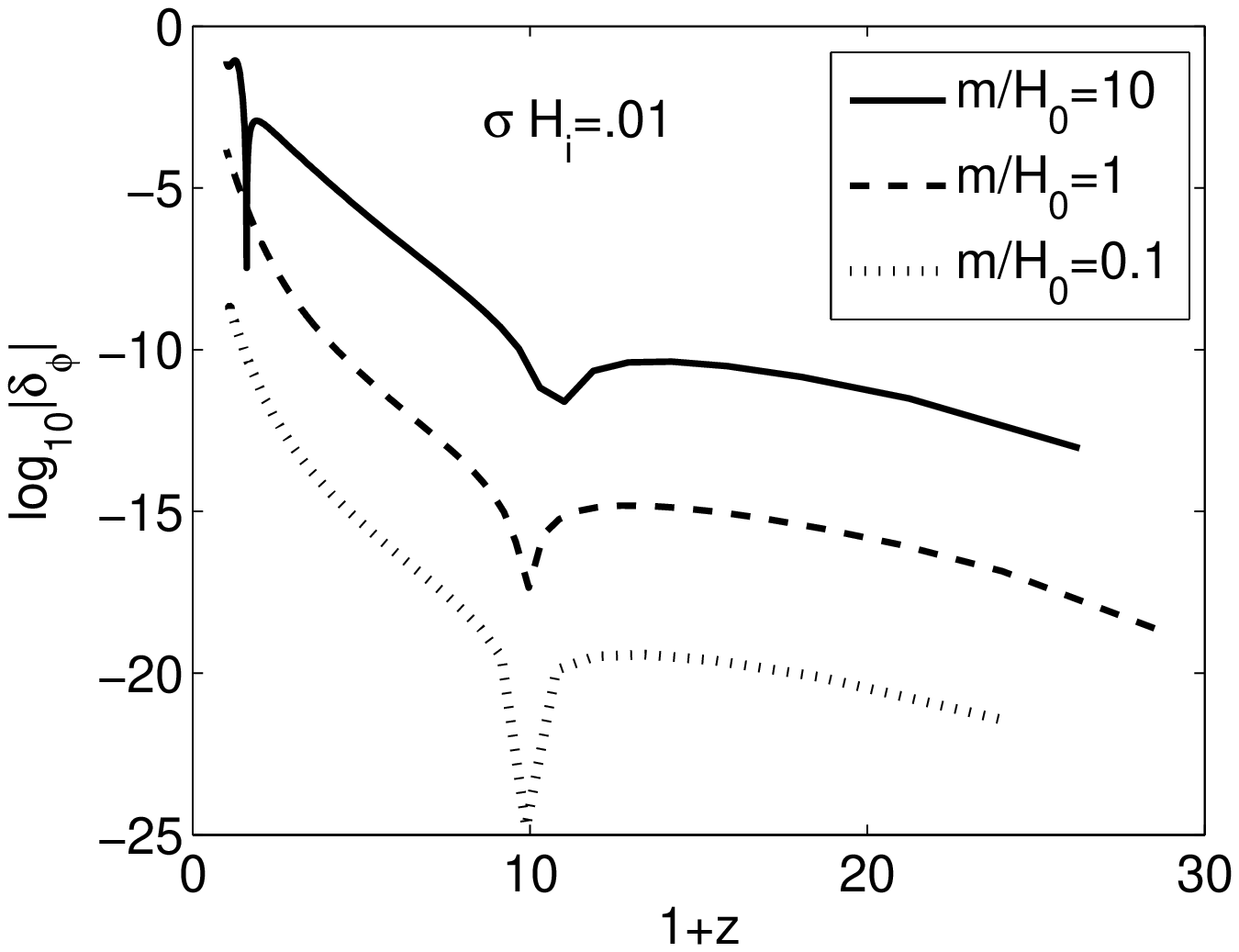,height=55mm}
	\caption{\label{mass_effect2}
    Same as Figure \ref{mass_effect1} with the y-axis on a
    logarithmic scale. The perturbation's is extremely
    sensitive to the mass scale.
}
\end{figure}

Figures \ref{mass_effect1} and \ref{mass_effect2} examine the sensitivity to the field's mass. As expected, increasing the mass of the scalar field causes $\delta_{\phi}$ to grow stronger and at increasingly earlier redshifts. One can also see that if the mass is heavy enough (an order of magnitude larger than the Hubble mass), the field will have had enough time to enter a period of rapid oscillations, and will effectively behave as regular matter.

\subsection{Equation of state}

\begin{figure}
	\epsfig{file=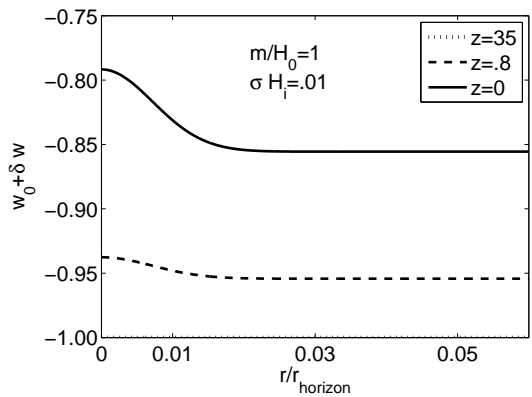,height=55mm}
	\caption{\label{total_w} Plot of $w_1$ vs $r$ for
	three different redshifts.
}
\end{figure}

\begin{figure}
	\epsfig{file=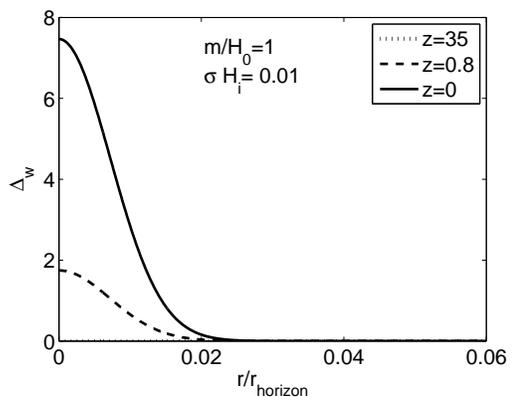,height=55mm}
	\caption{\label{Delta_w} Plot of $\%$ deviation in
	$w$ vs $r$ at three different redshifts.
}
\end{figure}

We would now like to focus on the local behavior of the equation of state (EOS) of the scalar field.
Let us define $w_0$ as the background homogeneous EOS, $\delta w$ as the leading order correction to $w_0$ and let the first-order corrected $w$ be
$w_1=w_0+\delta w$:
\begin{eqnarray}
	w_0 &=& \frac{p_0}{\rho_0} =
    \frac{\frac{1}{2}\dot\phi^2-V}
	{\frac{1}{2}\dot\phi^2+V} \\
	\delta w &=& \frac{1}{\rho_0}
    \left(\frac{}{}\delta p-w_0\delta\rho \frac{}{}\right) \\
	w_1 &=& w_0+\delta w ~,
\end{eqnarray}
where to first order $\delta \rho=\dot\phi\delta\dot\phi+\delta V$ and $\delta p = \dot\phi\delta\dot\phi-\delta V$, and we have suppressed the $\phi$ subscripts.

Figure \ref{total_w} shows how $w_1$ increases with time and becomes non-homogenous. The EOS at the perturbation is less negative than the background, but for this choice of mass it is still negative enough to behave as dark energy. \\

To quantify the extent of the inhomogeneity in $w$ we define $\Delta_w$ to characterize the \% deviation of the local $w$ from the background value,
\begin{eqnarray}
	\Delta_w = 100\bigg\vert\frac{\delta w}{w_0}
	\bigg\vert    
	= 100\bigg\vert \frac{1}{p_0}\left(\frac{}{}
    \delta p-w_0 \delta\rho \frac{}{}\right) \bigg\vert ~.
\end{eqnarray}
The evolution of the $\Delta_w$ spatial profile is shown in Figure \ref{Delta_w}.

\section{Discussion}
\label{Discussion}

\subsection{Void formation}

\begin{figure}
	\epsfig{file=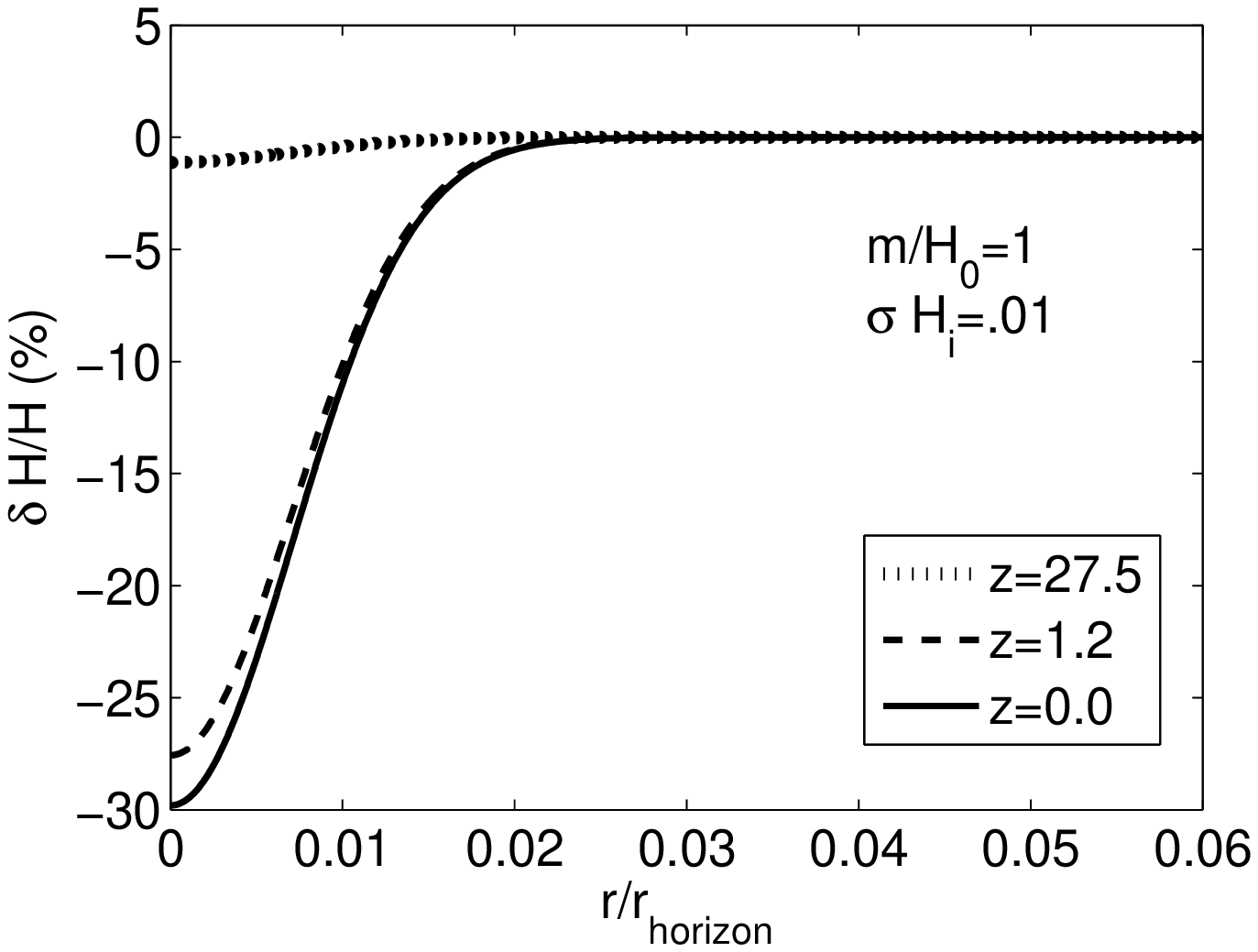,height=55mm}
	\caption{\label{dH_vs_r} Percentage variation of the local Hubble parameter at three different redshifts}
\end{figure}

\begin{figure}
	\epsfig{file=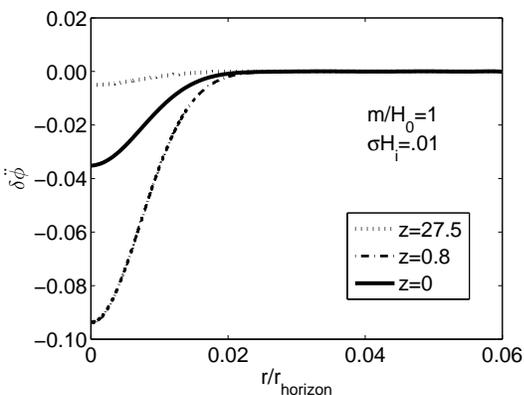,height=55mm}
	\caption{\label{de_accn_vs_r} Profile of $\delta\ddot\phi$ at three different redshifts}
\end{figure}

We now aim to intuitively explain why the DDE perturbation initially increases and then sharply drops in the presence of the matter perturbation. Initially, both the scalar field and the Hubble function are homogenous. The matter perturbation introduces an inhomogeneity in the Hubble function, as regions with matter overdensity expand slower due to the increased gravitational pull. As a result, $H$ acquires a spatial profile which evolves in time along with the matter overdensity, due to the $\rho\chi$ term in equation \eqref{drho1}. This is illustrated in Figure \ref{dH_vs_r}, which plots the spatial profile of the \% deviation of $H$ from the homogeneous background value, at three different redshifts. Regions which have a lower local value of $H$ offer less Hubble damping to the scalar field, as the $\dot\phi\chi$ contribution in equation \eqref{dphi1} is negative. Therefore, in these regions the scalar field accelerates down its potential slightly faster, the local $\ddot\phi+\delta\ddot{\phi}$ has a bigger absolute value than the acceleration of the background, $\ddot\phi$. Thus the matter perturbation imparts a local ``downhill kick'' to the field, and the strength of the kick depends on the magnitude of the matter perturbation at that point. The presence of matter is essential to trigger this mechanism. A similar conclusion was reached in \cite{mnf}.

Since the matter perturbation is gaussian, we can expect the profile of $\delta\ddot{\phi}$ to be gaussian as well, which is confirmed in figure \ref{de_accn_vs_r}. The acceleration of the field perturbation $\delta\ddot\phi$ is zero initially ($z=35$), but it quickly takes on a gaussian profile, which initially grows, and then shrinks at later times. The important point to note in figure \ref{de_accn_vs_r} is that the acceleration weakens, but doesn't change its sign throughout the evolution.

The acceleration profile leads to spatial variations in the energy density of the DDE. For the mass potential, the linear order of the density perturbation is $\delta\rho_\phi= \dot{\phi}\delta\dot{\phi} +m^2 \phi\delta\phi$, where the first term represents the local variation in kinetic energy (KE) and the second term represents the local variation in potential energy (PE). Whether the initial value of the scalar field was shifted to the right or left of its minimum, $\dot\phi$ and $\delta\dot\phi$ will be of the same sign, assuring a positive correction to the KE. On the other hand, $\phi$ and $\delta\phi$ will be of opposite signs, assuring the PE correction will be negative. The total correction to the energy density then will be positive when the KE dominates, creating an overdensity, and negative when the PE correction dominates, creating a void. The fact that $\delta\ddot\phi$  weakens but doesn't change its sign assures that $\delta\dot\phi$ approaches a constant value, but $\delta\phi$ keeps growing. The PE contribution
$m^2\phi\delta\phi$ soon becomes dominant over the KE correction, $\dot{\phi}\delta\dot{\phi}$.
As soon this happens, a void is created.\\
Similar reasoning should apply to other slow-roll potentials.

\subsection{Generality}

The results we have presented so far were for the mass potential, equation \eqref{vmass}. This is not a very attractive model from the theoretical point of view, as the choice of the mass scale and initial conditions are fine tuned.

While the general problem of fine tuning has not yet been resolved, the issue of the initial conditions is alleviated in tracking potentials such as equation \eqref{v2exp}. To verify that our results are not unique to the mass potential, we have repeated the analysis for the double exponential potential. We present here the equivalent of figures \ref{profiles} and \ref{total_w} to show that the results are essentially similar: as a reaction to the perturbation in the matter fluid, the DDE quickly forms a void. Starting very low, the rate of growth of its amplitude is significantly faster than that of the matter perturbation, motivating a further investigation into the non linear regime.

\begin{figure}
	\epsfig{file=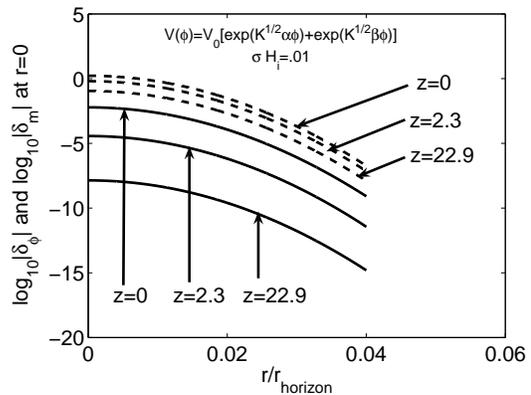,height=55mm}
	\caption{\label{fig12}
    The equivalent of figure \ref{profiles}, with the double
    exponential potential, equation \eqref{v2exp}.
}
\end{figure}
\begin{figure}
	\epsfig{file=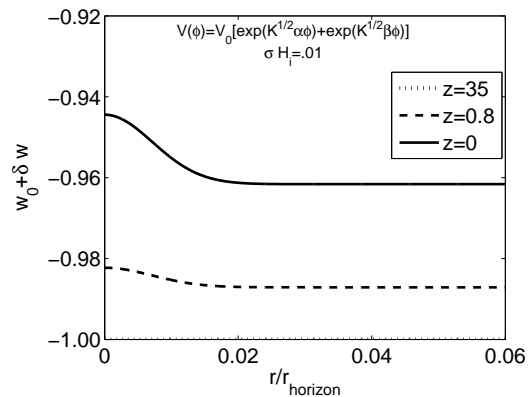,height=55mm}
    \caption{\label{fig13}
    The equivalent of figure \ref{total_w}, with the double
    exponential potential, equation \eqref{v2exp}.
}
\end{figure}

\section{Conclusions}
\label{Conclusions}

In this work we have investigated the clustering properties of DDE. We modeled the DDE as a scalar field with a light mass, and have shown that in the vicinity of gravitationally collapsing matter, the DDE develops inhomogeneities and forms voids. Our results show a high sensitivity to the mass scale of the field. For a mass much larger or smaller than the Hubble scale, the field imitates the behavior of dust or the cosmological constant, respectively. The interesting dynamics is most prominent in the window where the mass and Hubble scales are comparable. This window is within the relevant mass range for dark energy models. As the Hubble function was larger in the past, heavier fields would have had comparable mass and Hubble scales at some point in the past.

Our results should apply to any model of DDE which achieves the present acceleration through a slow roll phase, as the slow roll assures that only a small patch of the potential is probed. We have shown this explicitly for the double exponential potential. Whether our results apply to an even wider class of models which achieve accelerated expansion not through slow roll should be further investigated.

One thing which is clear from our results, is that DDE has potentially non-trivial behavior during the growth of inhomogeneities, though full non-linear analysis is needed to confirm whether the amplitude of the DDE inhomogeneities is relevant to observations. As inhomogeneities of dark energy are a clear signature differentiating between the cosmological constant and DDE, such possibilities should be fully explored and exhausted.

A full treatment of the observational consequences of our results is beyond the scope of this work, but we would like to mention a few possibilities. Some obvious places to look for DDE inhomogeneities include lensing, the ISW effect in the CMB, number counts, and mass functions. Some of these directions are being pursued \cite{parametric,cde}.

Our results show that both the energy density and the EOS of DDE develop
a spatial dependence. Thus any observation constraining either of the
above which can separately be measured locally and globally is valuable. \\

Another possibility which we would like to point out, is that it would
be useful to quantify the effect of a statistical distribution of DDE
voids on the CMB, or, following reference \cite{nhsn}, on the
directional distribution of supernovae. For example, detecting an
angular inhomogeneity in $H$ which is not in accordance with the matter
distribution might suggest the presence of a DDE void.

\acknowledgments{
Our sincere thanks to Andrew Liddle for reading the manuscript and pointing out some corrections, and mostly for the stimulating discussions which helped shape this work. We are also grateful to Tanmay Vachaspati for his careful reading of the manuscript, and for helpful feedback and discussions throughout the course of this work. We thank Daniele Steer and Alex Vilenkin for reading the manuscript, and Josh Frieman
for useful discussions. SD would like to thank the Institut d'Astrophysique, Paris, especially Daniele Steer and Patrick Peter, for their hospitality. The work of SD and IM is supported by the DOE and NASA.
}

{}

\end{document}